\documentclass{emulateapj}
\slugcomment{\today}
\def\arcdeg{\hbox{$^\circ$}}

\def\arcsec{\hbox{$^{\prime\prime}$}}

\def\snid{\ifmmode{\rm \tt SNID}\else{\tt SNID}\fi}
\def\dm15{\ifmmode{\Delta m_{15}}\else{$\Delta m_{15}$}\fi}
\def\magarcsec2{\ \rm{mag
\ arcsec}^{-2}}

\shorttitle{Light Echoes from Historic Supernovae}
\shortauthors{Rest et al.}

\begin{document}

\title{Scattered-Light Echoes from the Historical Galactic Supernovae Cassiopeia  A and Tycho (SN 1572)}

\author{A. Rest\altaffilmark{1,2,3}, D. L. Welch\altaffilmark{4},
N. B. Suntzeff\altaffilmark{5}, L. Oaster\altaffilmark{4},
H. Lanning\altaffilmark{6,14},K. Olsen\altaffilmark{6},
R. C. Smith\altaffilmark{1}, A. C. Becker\altaffilmark{7},
M. Bergmann\altaffilmark{8}, P. Challis\altaffilmark{9},
A. Clocchiatti\altaffilmark{10}, K. H. Cook\altaffilmark{11},
G. Damke\altaffilmark{1}, A. Garg\altaffilmark{2},
M. E. Huber\altaffilmark{11,12}, T. Matheson\altaffilmark{6},
D. Minniti\altaffilmark{10}, J. L. Prieto\altaffilmark{13},
W. M. Wood-Vasey\altaffilmark{9}
}

%A. Rest, D. L. Welch, N. B. Suntzeff, L. Oaster, H. Lanning,K. Olsen, R. C. Smith, A. C. Becker, M. Bergmann, P. Challis, A. Clocchiatti, K. H. Cook, G. Damke, A. Garg, M. E. Huber, T. Matheson, D. Minniti, J. L. Prieto, W. M. Wood-Vasey
% Cerro Tololo Inter-American Observatory, Harvard University, McMaster University, Texas A\&M University, National Optical Astronomy Observatory, University of Washington, Gemini Observatory, Harvard-Smithsonian Center for Astrophysics,Pontificia Universidad Cat\'olica de Chile, Lawrence Livermore National Laboratory, Johns Hopkins University, Ohio State University

\altaffiltext{1}{Cerro Tololo Inter-American Observatory (CTIO), Colina el Pino S/N, La Serena, Chile}

\altaffiltext{2}{Department of Physics, Harvard University, 17 Oxford Street, Cambridge, MA 02138}

\altaffiltext{3}{Goldberg Fellow}

\altaffiltext{4}{Dept. of Physics and Astronomy, McMaster University,
Hamilton, Ontario, L8S 4M1, Canada}

\altaffiltext{5}{Dept. of Physics, Texas A\&M University, College Station, TX 77843-4242}

\altaffiltext{6}{National Optical Astronomy Observatory, 950 N. Cherry Ave., Tucson, AZ 85719-4933}

\altaffiltext{7}{Dept. of Astronomy, University of Washington, Box 351580, Seattle, WA 98195}

\altaffiltext{8}{Gemini Observatory, Casilla 603, La Serena, Chile}

\altaffiltext{9}{Harvard-Smithsonian Center for Astrophysics, 60 Garden St., 
Cambridge, MA 02138.} 

\altaffiltext{10}{Dept. of Astronomy, Pontificia Universidad Cat\'olica de Chile, Casilla 306, Santiago 22, Chile}

\altaffiltext{11}{Lawrence Livermore National Laboratory, 7000 East Ave., Livermore, CA 94550}

\altaffiltext{12}{Johns Hopkins University, Baltimore, MD 21218}

\altaffiltext{13}{Dept. of Astronomy,
Ohio State University, 140 West 18th Ave., Columbus, OH 43210-1173}

\altaffiltext{14}{Deceased December 20th, 2007}

%%%%%%%%%%%%%%%%%%%%%%%
%%%%%%%%%%%%   ABSTRACT
%%%%%%%%%%%%%%%%%%%%%%%
 
\begin{abstract}
We report the discovery of an extensive system of scattered light echo
arclets associated with the recent supernovae in the local
neighbourhood of the Milky Way: Tycho (SN 1572) and Cassiopeia
A. Existing work suggests that the Tycho SN was a thermonuclear
explosion while the Cas~A supernova was a core collapse
explosion. Precise classifications according to modern nomenclature
require spectra of the outburst light. 
In the case of ancient SNe, this can only be done with spectroscopy of
their light echo, where the discovery of the light echoes from the
outburst light is the first step.
Adjacent light echo positions suggest
that Cas~A and Tycho may share common scattering dust structures. If
so, it is possible to measure precise distances between historical
Galactic supernovae.  On-going surveys that alert on the development
of bright scattered-light echo features have the potential to reveal
detailed spectroscopic information for many recent Galactic
supernovae, both directly visible and obscured by dust in the Galactic
plane.
\end{abstract}


\section{Introduction}
\label{sec:intro}

The suggestion that historical supernovae might be studied by their
scattered light echoes was first made by \cite{Zwicky40} and attempted
by \cite{vandenBergh65,vandenBergh75}. Our group pioneered the
discovery and study of ancient supernova scattered-light echoes using
difference imaging in the LMC field where three echo complexes were
found to be associated with 400-900 year-old supernova remnants (SNR)
\citep{Rest05b}. We have since obtained a spectrum of one of these
echoes which reveals that the echo light is from the class of
over-luminous Type Ia supernovae \citep{Rest08,Badenes08} and
demonstrating that precise modern supernova classifications are
possible for ancient supernovae. These scattered light echoes
preserve optical spectral line information from the outburst, and will
be useful for future spectroscopic studies of the original SN
light. This is in contrast to the moving Cas~A features (sometimes
called “infrared echoes”) identified using far-infrared imagery from
the Spitzer Space Telescope \citep{Krause05}, which are the result of
dust absorbing the outburst light, warming and re-radiating at longer
wavelengths.

Echo features similar to those found in the LMC should be detectable
within our own Milky Way. The challenge has been to locate them across
a much larger solid angle. We have begun a program to find echoes
around a sample of 7 certain historical Galactic supernovae recorded
in the last 2000 years \citep{Stephenson02} (SN 185 AD/Centaurus, SN
1054 AD/Crab, SN1006 AD/Lupus, SN 1181 AD/Cassiopeia, Tycho, Kepler,
Cas~A). Given the well-constrained ages of these historical supernovae
and estimated distances, we can improve our chance to find echoes by
targeting regions of cold dust at the approximate expected angular
distance. We used the reprocessed 100 $\micron$ IRAS images
\citep{Miville05} to select fields with lines of sight which contain
such dust, choosing fields closer to the Galactic plane than the
supernovae in the expectation that dust would be more highly
concentrated there.

Tycho in 1572 discovered one of the last two naked eye SNe in the
Galaxy, while another nearby supernova, Cassiopeia A, evidently
escaped discovery around 1671 \citep{Stephenson02}. Based on the
properties of the associated supernova remnants, it is thought that
the Tycho SN was a thermonuclear explosion \citep{Ruiz04a,Badenes06}
while the Cas~A supernova was a core collapse explosion
\citep{Chevalier78}.

%%%%%%%%%%%%%%%%%%%%%%
%%%%%%%%%%%%   OBSERVATIONS
%%%%%%%%%%%%%%%%%%%%%%
\section{Observations \& Reductions}
\label{sec:observations}

We obtained images from four observing runs on the Mayall 4m telescope
at Kitt Peak National Observatory in the fall of 2006 and 2007.  The
Mosaic imager, which operates at the f/3.1 prime focus at an effective
focal ratio of f/2.9, was used with the Bernstein VR Broad filter
(k1040) which has a central wavelength of 594.5nm and a FWHM of
212.0nm. Images of target fields in Cassiopeia were obtained on UT
2006 Oct 21-23, 2006 Dec 16, 2007 Oct 12-15, and 2007 Dec 13-15.
% (see Table~\ref{tab:observations}). 
Exposure times were between 120 and 150 seconds. The interval between 
the two epochs for a given field is at least 53 days and as much as two 
years. We expect echo arclets to have typical apparent motions of 20-40 
arcsec yr$^{-1}$, meaning a 3-month baseline is sufficient to resolve 
their apparent motion.  Imaging data was kernel- and flux-matched, 
aligned, subtracted, and masked using the SMSN
pipeline \citep{Rest05a,Garg07,Miknaitis07}. The resulting difference
images are remarkably clean of the (constant) stellar background and
are ideal for searching for variable sources.

\section{Analysis}
\label{sec:analysis}

\begin{figure}[t]
\epsscale{0.9}
\plotone{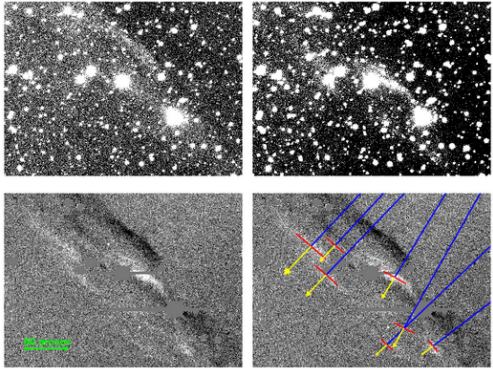}
\caption[]{
Light echo arclets associated with Tycho from field \#4821. 
The orientation is N up and E to the left and the images are 
325 x 250 arcseconds. The upper two panels show the first epoch 
image from 20 October 2006 (left), and
the second epoch image from 13 December 2007 (right). The lower images
are the difference images between the two upper images where white
represents the later (October 2006) image and black the earlier
December image. Saturated bright stars are masked gray. 
In the lower panel, the left image is repeated in the right
panel with the motion vectors plotted. Red represents a straight line
fit to the arclet, yellow represents the apparent motion of the
arclet, and blue shows the reverse vector direction. 
The VR surface brightness in the brighter arclets is roughly
$~24 \magarcsec2$. The widths of the echoes are resolved, and typically
$~10\arcsec$ across.
\label{fig:leexample}}
\end{figure}

Using the same techniques developed for the LMC echo
searches \citep{Rest05b}, candidate echo arclets, such as those shown
in Figure~\ref{fig:leexample}, were identified by visual examination
of difference images.
% (see Table~\ref{tab:le}). 
%

\begin{figure}[t]
\epsscale{1.0}
%\ifsubmode
\plotone{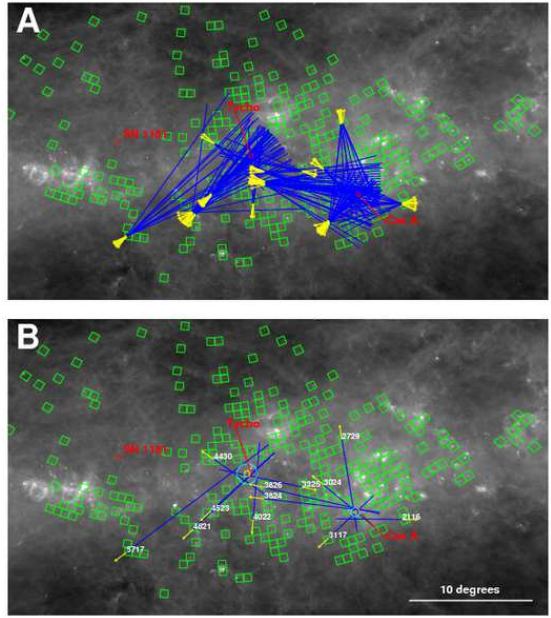}
%\fi
\caption[]{
Arclet vector motions in the region of Cas~A and Tycho supernovae. The
vectors are plotted on an image reconstructed from IRAS data at
$100 \micron$ \citep{Miville05}. The panels are $44\arcdeg \times
25\arcdeg$ with N up and E to the right. The green squares show the
Mosaic fields where we have at least two epochs and have been searched
for arcs in the difference image. The red circles mark the positions
of the three historic SNe SN 1181, Tycho, and Cas~A. There are six
clusters of light echoes with apparent motion vectors pointing back to
the Cas~A SNR, and six more consistent with an origin coincident with
the Tycho SNR. Panel B shows the average vector for each light echo
cluster. The apparent positions for the points of origin for the two
echo complexes are listed in Table~\ref{tab:results}, and calculated
as average of the all the pairs of vector crossings (clipped at
3-sigma) where the large light blue circles denote the standard
deviation of the crossings of all vector pairs in each echo
complex. The yellow circles are centered on the
Table~\ref{tab:results} mean vector crossings and the circle sizes
show the error in the mean for the vector crossings. 
\label{fig:vectors}}
\end{figure}

We estimated the arc motion directions by eye and plotted the inverse
motion vectors, as shown in panel {\it A} of Figure~\ref{fig:vectors}.
Two echo complexes were discovered. In the first, we found six
clusters of light echoes with proper motion vectors converging back to
the Cas~A SNR, and in the other, six more echo clusters consistent
with an origin coincident with the Tycho SNR. No echo arclets were
detected for SN 1181 during this search, which also lies within our
search area, but in a region of lower 100 micron surface
brightness. All light echo features discovered seem to be associated
with either Cas~A or Tycho. We have obtained 3rd and 4th epoch images
in 2007 for the light echo groups we detected in 2006, and the light
echoes were redetected in these images.

%(see column PA(stdev) in
%Table~\ref{tab:leaverage}) 
For a given light echo cluster, the vectors have a spread in angle of
~10 degrees (see column PA(stdev) in Table~\ref{tab:leaverage}) mainly due to the orientation of the scattering dust: If
the reflecting dustsheet or filament is confined to a plane
perpendicular to the line of sight, the light echo vector points
exactly back to the source. However, if the dustsheet is inclined or
warped, then the tangent to the light echo arc may rotate with respect
to the perpendicular direction to the remnant position.  Provided that
the inclinations of dust filaments in azimuth and in distance are not
correlated, the average vector will still point in the direction of
the SNR.  For each light echo cluster, we calculate the average
vector
(see Table~\ref{tab:leaverage}),
as shown in panel {\it
B} of Figure~\ref{fig:vectors}.  The estimated positions for the
points of origin for the two echo complexes are given in
Table~\ref{tab:results} as calculated by the average of all pairs of
vector crossings (clipped at 3-sigma). The Tycho and Cas~A SNR
positions are within the standard deviation of the points of origin
and within 3-sigma of the average position.
%
%\begin{deluxetable}{llllllllllllll}
%\begin{deluxetable}{llllrrrrrrllll}
\begin{deluxetable*}{llllrrrrrrllll}
\tabletypesize{\tiny}
\tablecaption{
\label{tab:leaverage}}
\tablehead{
\colhead{Field\#} & \colhead{SNR} & \colhead{RA} & \colhead{Dec} & \colhead{MJD} & \colhead{N} & \colhead{PA} & \colhead{PA(stdev)} & \colhead{DS} & \colhead{$z$} & \colhead{RAmin} & \colhead{Decmin} & \colhead{RAmax} & \colhead{Decmax}\\
\colhead{(0)} & \colhead{(1)} & \colhead{(2)} & \colhead{(3)} & \colhead{(4)} & \colhead{(5)} & \colhead{(6)} & \colhead{(7)} & \colhead{(8)} & \colhead{(9)} & \colhead{(10)} & \colhead{(11)} & \colhead{(12)} & \colhead{(13)}
}
\startdata
 2116 &  CasA & 23:02:42.9 & +56:48:18 &  66.46 &  12 & -117.4 & 16.4 &  3.4 &  431.9 & 23:02:37.7 & +56:44:11 & 23:02:53.0 & +56:50:50\\
 2729 &  CasA & 23:13:36.6 & +64:41:15 & 130.77 &   9 &  -15.6 & 11.1 &  6.0 & 1374.7 & 23:12:53.9 & +64:40:25 & 23:14:48.1 & +64:42:37\\
 3024 &  CasA & 23:37:53.6 & +61:42:55 & 116.49 &   6 &   31.0 &  9.5 &  3.4 &  432.3 & 23:37:50.9 & +61:42:27 & 23:37:57.2 & +61:43:17\\
 3117 &  CasA & 23:45:30.9 & +57:26:51 &  97.86 &  22 &  117.3 & 15.0 &  3.2 &  373.6 & 23:45:11.0 & +57:20:50 & 23:46:13.8 & +57:35:10\\
 3824 &  CasA & 00:19:03.0 & +61:45:46 &  87.87 &  37 &   73.3 & 11.4 &  7.5 & 1966.7 & 00:18:02.7 & +61:36:10 & 00:19:36.5 & +61:55:45\\
 3826 &  CasA & 00:17:39.7 & +62:40:59 &  86.44 &   9 &   66.7 & 10.1 &  7.7 & 2034.5 & 00:17:29.5 & +62:39:48 & 00:17:45.1 & +62:44:06\\
 3325 & Tycho & 23:52:04.4 & +62:03:19 & 170.51 &   6 & -118.2 &  9.1 &  4.3 &  133.9 & 23:52:01.6 & +62:02:35 & 23:52:07.8 & +62:04:20\\
 4022 & Tycho & 00:28:25.6 & +60:10:14 & 121.57 &   5 &  168.0 &  4.8 &  4.0 &   93.7 & 00:28:20.8 & +60:10:03 & 00:28:30.7 & +60:10:29\\
 4430 & Tycho & 00:52:10.1 & +65:28:54 & 144.70 &   4 &   52.3 &  7.8 &  3.2 &  -18.8 & 00:52:08.5 & +65:28:50 & 00:52:11.3 & +65:28:60\\
 4523 & Tycho & 00:55:27.1 & +61:10:13 & 227.40 &  17 &  132.2 &  9.0 &  4.6 &  179.4 & 00:55:19.7 & +61:05:08 & 00:55:35.6 & +61:14:58\\
 4821 & Tycho & 01:07:11.6 & +59:38:37 & 243.77 &  21 &  135.5 & 17.1 &  6.7 &  547.8 & 01:05:48.7 & +59:31:31 & 01:08:21.8 & +59:44:14\\
 5717 & Tycho & 01:46:38.0 & +57:13:36 & 216.84 &  10 &  138.0 &  9.3 & 12.1 & 1607.1 & 01:46:29.0 & +57:12:42 & 01:46:50.1 & +57:14:15\\
\enddata
\tablecomments{
This Table lists the parameters of the average light echo arclets and
their associated vectors.  The column Field\# shows the number of the
field in which the light echo was found. The column SNR indicates the
SNR associated with the light echo. RA, Dec define the base position
of the vector at position angle of PA in degrees. MJD indicates the
average modified Julian date MJD - 54000 of the observations.  The
column N shows how many light echo arclets where used.  The column
PA(stdev) is the standard deviation of the position angles of all vectors
in a given light echo cluster, 3-sigma clipped in order to
remove outliers.  Column DS shows the angular separation between the
light echo and the associated SNR in degrees. The inferred distance $z$ in light
years of the dust along the line of sight from the SNR is shown in
column $z$. The light echoes of a given light echo cluster are within
the box specified by RAmin, DECmin, RAmax, DECmax. All positions are equinox
J2000.0
}
%\end{deluxetable}
\end{deluxetable*}

%\begin{deluxetable*}{ccccccc}
\begin{deluxetable}{cccccc}
\tabletypesize{\scriptsize}
\tablecaption{
\label{tab:results}}
\tablehead{
\colhead{SNR} & \colhead{RA(SNR)} & \colhead{Dec(SNR)} & \colhead{RA(origin)} & \colhead{Dec(origin)} & \colhead{$\delta r$}\\
\colhead{(0)} & \colhead{(1)} & \colhead{(2)} & \colhead{(3)} & \colhead{(4)} & \colhead{(5)}
}
\startdata
  Cas A & 23:23:24 & +58:48:54 & 23:25:16 & +58:46:43 &          6.2 \\
 Tycho & 00:25:08 & +64:09:56 & 00:27:53  & +64:03:25 &         14.5 \\
\enddata
\tablecomments{
RA(SNR) and Dec(SNR) are the radio positions of the SNR likely
associated with the light echo group. RA(origin) and Dec(origin) are
the averaged position of the vector crossing points. The uncertainties
$\delta r$ in the position are given in arcminutes. Coordinates are
equinox J2000.
}
%\end{deluxetable*}
\end{deluxetable}

%

%%%%%%%%%%%%%%%%%%%%%%%%%%%%%%%%%%%%%

\section{Discussion \& Conclusions}
\label{sec:discussion}

\begin{figure}[t]
\epsscale{0.85}
%\ifsubmode
\plotone{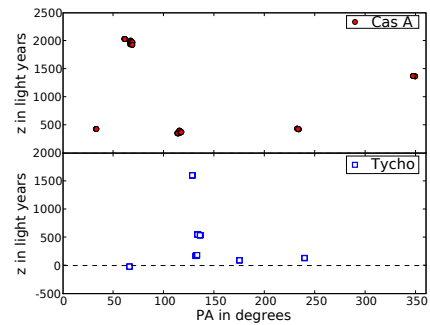}
%\fi
\caption[]{
The distance $z$ from the supernova to the scattering cloud projected
along our line-of-sight plotted with respect to the position angle
from the associated supernova remnant. To calculate $z$, we assume that
the distance to Tycho is 2300pc and for Cas~A
3400pc.
\label{fig:z_vs_PA}}
\end{figure}

The light echo equation \citep{Couderc39} 
\begin{equation}
z = \frac{\rho^2}{2ct} - \frac{ct}{2}
\end{equation}
relates the depth coordinate, $z$, the echo-supernova distance
projected along the line-of-sight, to the echo distance $\rho$
perpendicular to the line of sight, and the time $t$ since the
explosion was observed.  Then the distance $r$ from the scattering
dust to the SN is $r^2=\rho^2+z^2$, and $\rho$ can be estimated with
$\rho \approx D \sin\alpha$, where $D$ is the distance from the
observer to the SN, and $\alpha$ is the angular separation between the
SN and the scattering dust, which yields the 3-D position of the dust
associated with the arclet. In Figure~\ref{fig:z_vs_PA} we see that
three of the six Cas~A light echo clusters (field \#2116, \#3024, and
\#3117) are at very similar distances at $z \approx 400$ light years in front of
Cas~A.  This clustering in $z$ suggests that these three clusters are
associated with the same extended dust sheet/filament. Similarly, two
other Cas~A light echo clusters (field \#3824 and \#3826) are likely
from a single dust structure at $z \approx 2000$ light years. For Tycho a
number of arclets (fields \#3325, \#4022, \#4430, and \#4523) are found
near $z \approx 100$ light years (relative to the Tycho SNR) and one field
\#5717 contains scattering dust at $z \approx 1600$ light years.

% AR
The distance to Tycho and Cas A is estimated to be $2.3 \pm 0.5$kpc
and $3.4 \pm 0.5$kpc, respectively
\citep{Albinson86,Strom88,Lee04,Reed95}. This implies that Cas A is
$3600 \pm 2300$ light years behind Tycho.  For both Cas A and Tycho,
there is a clustering of the scattering dust structures in $z$,
indicating that there are extended dust structures or sheets, in
contrast to small, local dust structures.  The Tycho scattering dust
structures cluster around $z=0$, implying that there are also in fron
of Cas A. Since Cas A is in close angular proximity to Tycho, it can
be expected that dust belonging to this extended dust structure produce
light echoes for both Tycho and Cas A. Thus 
% AR end
we explore the possibility that Cas~A and Tycho share extended
scattering dust structures. It is notable that the difference between
the outer and the inner dust structures is about 1500-1600 light years
for both Cas~A and Tycho. Thus one scenario is that the Tycho light
echo groups in field \#3325, \#4022, \#4430, and \#4523 and Cas~A
light echo groups in field \#2116, \#3024, and \#3117 belong to the
same dust structure, and Tycho \#5717 and Cas~A \#3824 and \#3826 to
another dust structure. Since the measured $z$ distance is always
relative to its associated SNR, this would then imply that Cas~A is
about 300-400 light years farther away than Tycho, which 
% AR
is not within 1 sigma
of the current estimate of $3600 \pm 2300$ light
years. 
% AR end
Another possibility is that the outer Cas~A light echo cluster
in fields \#3824 and \#3826 arise from the same extended dust
structure which causes the Tycho light echoes in field \#3325, \#4022,
\#4430, and \#4523. This association seems more likely due to the
close proximity of these two arc groups on the sky.  For this
association, Cas~A must be $~1900$ light years more distant than
Tycho, 
% AR
which is within the errors of the current estimates.
% AR end

Geometric light echo distance estimates are possible when the
scattering angle is known (e.g. \cite{Sparks08} for V838 Mon). A
maximum in the linear polarization is expected for an angle of $90
\arcdeg$ - when the scattering dust is at the same distance as the
supernova. Therefore this method works best with light echoes with 
scattering angles spread around $90 \arcdeg$, which corresponds to 
distances spread around $z=0$. Such a situation may exist for Tycho
(see Figure~\ref{fig:z_vs_PA}). In the worst case, the distance
$D_{tycho}$ may be significantly larger than the current estimate of
2300pc \citep{Albinson86,Strom88,Lee04} and a strong lower limit on the distance of Tycho
may be set. If Tycho is nearer than the current estimate, a more
accurate distance is possible since the distance of the scattering
dust will then bracket the supernova distance. For Cas A, none of our
detections imply scattering dust behind the SNR (see
Figure~\ref{fig:z_vs_PA}).  However, the existence of re-radiated echo
light in the infrared \citep{Krause05} suggests that such echoes may
yet be detected.

\begin{figure}[t]
\epsscale{0.85}
%\ifsubmode
\plotone{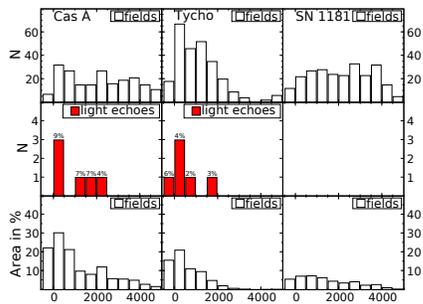}
%\fi
\caption[]{
Histograms of the number of fields with two epochs versus their
distance z along the line of sight (i.e., how much “in front” or “in
back” of the supernova the scattering dust is) to the SNRs Cas~A,
Tycho, and SN 1181 from left to right in the upper panels.  The middle
panels show the number of light echo groups found in a given $z$ bin.
The percentage of fields with light echoes for a given bin is shown
above the red bars in percent. The bottom panels show the \% of the
total area (i.e. the annulus associated with the given $z$ bin) is
covered by the observed fields. 
\label{fig:histo}}
\end{figure}

For Cas~A and Tycho, 5\% and 3\% of the fields surveyed with $z<2000$
light years contain scattered-light echo arclets, respectively (see
upper and middle panels of Figure~\ref{fig:histo}).  Virtually all
light echoes are found at $z<2000$ light years, and in particular at
$0<z<500$ light years.  No light echoes from SN1181 (3C58) were
detected. On first glance this is surprising.  However, the bottom
panels reveal that we have searched a smaller fraction of the $0<z<500$
light year, forward-scattering region of SN1181 in comparison to Cas A
and Tycho.  A combination of fainter surface brightness of the light
echoes due to the supernova age and brightness and small number
statistics might explain the lack of detections. Deeper surveys of
SN1181 region might yet yield light echoes.
Light echoes are expected to be scattered light of the averaged flux
around maximum luminosity - the only significant modification of the
event outburst light is expected to be due to dust grain size which,
in the absence of additional absorption along the echo paths, makes
the spectral flux bluer. We have demonstrated that the type of a SN
(Ia, II, etc.) and the subtype (luminous, normal, underluminous Type
Ia) can be determined centuries after the event by taking a spectrum
of the light echo \citep{Rest05b,Rest08}. 

%%%% beginning
%Prior to this work, the best technique for typing historical SNe came
%from radio and X-ray studies of the associated remnants.  The first
%proposed association for Tycho's SN outburst \citep{Hanbury52} was
%with a radio source at this position \citep{Bolton48}. Minkowski was
%reported to have discovered a ``faint gaseous filament'' lying close
%to the radio position and ``moving away from Tycho Brahe's position''
%\citep{Baldwin57}. Later observations confirmed the motion of the
%filaments associated with the remnant
%\citep{vandenBergh71,vandenBergh73}.  Modern X-ray spectra of the
%remnant have been interpreted by \cite{Ruiz04b,Ruiz04a,Badenes08} to
%be most consistent with a delayed-detonation model of a normal Type Ia
%SN.

%The nature of Cas A outburst  has proven much more elusive since 
%the outburst itself seems to avoided detection, despite having 
%happened in an era when observation of the sky was common 
%\citep{Stephenson02}.  A very bright
%and localized radio source \citep{Ryle48} in the vicinity of a star
%position measured by Flamsteed (but with no counterpart in the sky in
%modern times) was once considered to be an attractive candidate for the SN
%but the association is now generally discounted \citep{Stephenson02}. 
%We use the outburst date $1671.3 \pm 0.9$ determined from proper motions 
%of shell and ejecta knots in the Cas~A SNR \citep{Thorstensen01}. 
%Ground-based spectroscopy and spectral classification of supernova
%type is possible when the surface brightness of the echo light is brighter
%than $~22  \magarcsec2$.
%%%% end

In the Milky Way Galaxy alone there have been at least 7 historic SNe
\citep{Stephenson02} that are good candidates to have produced
still-observable light echoes. Other apparently young SNRs identified
in radio and x-ray surveys exist in the plane of the Galaxy
\citep{Green84,Reynolds08} that have no historical records, most
likely because they are obscured by dust. Light echoes of these
obscured SNe may be visible since the line-of-sight to scattering dust
may be less obscured than direct light. Potentially, dozens of ancient
SNe can be typed by the means of light echo spectroscopy as described
in \cite{Rest08}.  We note that this is one of the very rare occasions
in astronomy that cause and effect of the same astronomical event can
be observed, in that we can study the physics of the SNR as it appears
now and also the physics of the explosion which produced the SNR as it
appeared hundreds of years ago.

The study of scattered-light echoes from galactic SNe provide a host
of newly-recognized observational benefits which have only just begun
to be exploited including: the capacity to understand the connection
between remnant properties and the outburst spectral type, access to
observables related to asymmetric explosion properties, and a network
of absolute distance differences. 

%\section{Acknowledgments}
\acknowledgements
We dedicate this paper to Howard Lanning, who passed away December 20,
2007. AR thanks the Goldberg Fellowship Program for its support. DW
acknowledges support from the Natural Sciences and Engineering
Research Council of Canada (NSERC). SuperMACHO was supported by the HST
grant GO-10583 and GO-10903. This work was partially performed under
the auspices of the U.S.  Department of Energy by Lawrence Livermore
National Laboratory in part under Contract W-7405-Eng-48 and in part
under Contract DE-AC52-07NA27344. NBS thanks S. van den Bergh for
suggesting in 1995 to search for echoes using CCDs. NOAO is
operated by AURA under cooperative agreement with the NSF.

%%%%%%%%%%%%%%%%%%%%%%
%%%%%%%%%%%%   BIB
%%%%%%%%%%%%%%%%%%%%%%

\bibliographystyle{apj}
\bibliography{ms}

%%%%%%%%%%%%%%%%%%%%%%
%%%%%%%%%%%%   TABLES
%%%%%%%%%%%%%%%%%%%%%%
%\clearpage
%\input{tab1}
%\input{tab2}
%\input{tab3}
%\input{tab4}

\end{document}